\documentstyle[preprint,eqsecnum,aps]{revtex}

\begin{document}
\draft

\title{Variational calculations of the $\Lambda$-seperation energy of the
$_{\,\,\Lambda}^{17}$O hypernucleus}

\author{A.A.\ Usmani\cite{byline}}
\address{Interdisciplinary Laboratory, SISSA, I-34014,Trieste, Italy}
\author{Steven C. \ Pieper }
\address{Physics Division, Argonne National Laboratory, Argonne, IL 60439-4843}
\author{Q.N. \ Usmani}
\address{Department of Physics, Jamia Millia Islamia, New Delhi-110025, India}

\date{January 24, 1994}
\maketitle

\begin{abstract}
Variational Monte Carlo calculations have been made for the
$_{\,\,\Lambda}^{17}$O
 hypernucleus using realistic two- and three-baryon interactions. A two pion
exchange potential
with spin- and space-exchange components is used for the $\Lambda$N potential.
Three-body
 two-pion exchange and strongly repulsive dispersive $\Lambda$NN interactions
are also
 included. The trial wave function is constructed from pair- and
triplet-correlation
operators acting on a single particle determinant. These operators consist of
central,
spin, isospin, tensor and three- baryon potential components. A cluster Monte
Carlo
method is developed for noncentral correlations and is used with up to
four-baryon
clusters in our calculations. The three-baryon $\Lambda$NN force is discussed.

\end{abstract}


\newpage
\section{INTRODUCTION}
In this paper we initiate a variational Monte Carlo study of hypernuclei
using realistic two- and three-baryon interactions involving a $\Lambda$ and
nucleons. In
the past, variational calculations of the s-shell
hypernuclei,\cite{dalitz72,bodmer88}
a few p-shell hypernuclei using appropriate models\cite{bodmer84} and $\Lambda$
binding
to nuclear matter\cite{bodmer88,qusmani80} were performed using mostly
simplified and
central nucleon-nucleon (NN)
interactions. The aims of these calculations have largely been to deduce
information
about the $\Lambda$-nucleon ($\Lambda$N) and $\Lambda$-nucleon-nucleon
($\Lambda$NN)
interactions. In addition, as a result of these studies, one could also explore
the
structure of hypernuclei.

The reason for using simplified NN interactions is the hope that the
uncertainties in the
NN interaction will largely cancel out in these calculations. This is because
the
$\Lambda$-separation energies $B_{\Lambda}$, which are the main experimental
input in
these calculations, are the differences of the energies of hypernuclei and
their cores,
i.e., -- $B_{\Lambda}$ = $_{\Lambda}^{A}E$ -- $^{A-1}E$, where
$_{\Lambda}^{A}E$ is the total
energy  of the hypernucleus and $^{A-1}E$ is the ground state energy of the
core nucleus.
However, interactions generate strong NN correlations in the nuclear wave
function.
A realistic NN interaction will generate central, spin,
spin-isospin, tensor and other two-nucleon
correlations\cite{wiringa91,pandharipande79}.
In addition, there are significant three-nucleon correlations. In a
hypernucleus,
because of the operator dependence, these
correlations may interact in a complicated manner with the $\Lambda$N and
${\Lambda NN}$
correlations. This whole group of correlations then interact with the two- and
three-body
operators of the two- and three-baryon interactions. This can result in
important
contributions to the $\Lambda$-binding energy compared to the use of only
central NN correlations generated by purely central NN interactions.

In this study, we use realistic two- and three-nucleon interactions and wave
functions to
see their effects on hypernuclei. There is another important aspect of the
present study,
the development of a methodology and variational program for hypernuclei. For
the
p-shell hypernuclei,
we adopt the cluster Monte Carlo(CMC) technique developed in
Ref.\cite{pieper92} which we
refer to as PWP. As a first step, we study $_{\,\,\Lambda}^{17}$O.
Generalization to
$_{\,\,\Lambda}^{16}$O will be straightforward. Further development of the
one-body
part of the nuclear wave functions will be needed for other p-shell
hypernuclei. We
intend to cover a wide range of hypernuclei in order to have reliable
information on
three-baryon forces, because in this study we find that the role of the
three-body
$\Lambda$NN interaction is greatly altered from that found in some previous
studies.

There are a few calculations of hypernuclei in which realistic NN forces
have been used. One such calculation is by J. Carlson\cite{carlson91} in which
he explicitly
considers the $\Lambda$N $\rightarrow \Sigma$N channel in $_{\Lambda}^{5}$He
and
$_{\Lambda}^{4}$He using the Nijmegen interaction. This study shows that the
Nijmegen
interaction underbinds the four-body hypernuclei and that the five-body
hypernucleus is
unbound relative to a separated $\alpha$ and $\Lambda$ particle. Also, it does
not
reproduce the spin-splitting in the four-body hypernuclei. To resolve the
classical
over-binding\cite{dalitz72,gal75,povh80} problem of $_{\Lambda}^{5}$He, Bando
and
Shimodaya\cite{bando80},
and Shinmura $et\,al.$\cite{shinmura84} have also performed calculations with
Reid soft core and  Hamada-Johnston NN potentials by calculating effective
interactions
using a G-matrix approach. All these calculations of the $\Lambda$-seperation
energies
$B_{\Lambda}$ are confined to s-shell hypernuclei.

Since no experimental data for
$_{\,\,\Lambda}^{17}$O exists, we generate "pseudo-experimental" or
semi-empirical data for
this system. This is done in Sec. II where we also briefly discuss the
experimental status
of $\Lambda$-seperation energy data. In Sec. III we review the $\Lambda$ and
nucleon
two- and three-body potentials. Section IV deals with the variational wave
function. In
Sec. V we describe the techniques of our calculations for
$_{\,\,\Lambda}^{17}$O.
In Sec. VI we present our results, and Sec. VII contains our conclusions.

\section{$\Lambda$-separation energies}

If we combine the results of previous experiments summarized in
Ref.\cite{davis86},
the in-flight reaction (K$^{-},\pi^{-}$),\cite{milner85,pile91,chrien80} and
the
associated production\cite{bertini80} ($\pi^{+}$,K$^{+}$), we have now almost
30 well
established hypernuclei with a wide range of baryon numbers $A\le81$ and
orbital angular
momentum $\ell_{\Lambda}\le3$. The hypernuclei that are relevant for the
empirical
determination of the $B_{\Lambda}$ value of $_{\,\,\Lambda}^{17}$O are
$_{\,\,\Lambda}^{11}$C,
$_{\,\,\Lambda}^{12}$C, $_{\,\,\Lambda}^{13}$C, $_{\,\,\Lambda}^{16}$O,
$_{\,\,\Lambda}^{28}$Si,
$_{\,\,\Lambda}^{32}$S, $_{\,\,\Lambda}^{40}$Ca, $_{\,\,\Lambda}^{51}$V,
and $_{\,\,\Lambda}^{89}$Y.

We use here three approaches for the empirical $B_{\Lambda}$ value of
$_{\,\,\Lambda}^{17}$O.
In the first approach\cite{qusmani} microscopic calculations of $B_{\Lambda}$
of the above
hypernuclei were carried out with phenomenological two- and three-body
$\Lambda$N and
$\Lambda$NN interactions that were previously obtained from studies of
$\Lambda$p
scattering,
the s-shell hypernuclei, $_{\Lambda}^{9}$Be as a representative of p-shell
hypernuclei in
the 2$\alpha +\Lambda$ model, and the $\Lambda$-binding to nuclear
matter\cite{bodmer88,bodmer84,qusmani80}. The $\Lambda$-seperation energies
$B_{\Lambda}$
are obtained from a Schroedinger
equation with a $\Lambda$-nucleus potential $U_{\Lambda}$ and an effective mass
$m_{\Lambda}^{*}$ which are obtained in the local-density approximation using
the Fermi
hypernetted chain technique for the $\Lambda$ binding to nuclear matter. Such a
procedure
gives a good account of the $B_{\Lambda}$ data of the above hypernuclei for
$\ell_{\Lambda}\le3$.  Using this procedure, we calculate the difference,
$\Delta B_{\Lambda}$, of the $\Lambda$-separation
energies  of $_{\,\,\Lambda}^{17}$O and $_{\,\,\Lambda}^{16}$O.
This gives $\Delta B_{\Lambda} = 0.30$ MeV.

In the second approach, we use the purely phenomenological technique adopted by
Millener {\it et al.}\cite{millener88}. Here we use a Woods-Saxon
$\Lambda$-nucleus potential
whose parameters were fitted to the $B_{\Lambda}$ data of the above mentioned
hypernuclei,

\begin{equation}
U_{\Lambda}(r)={V_{0\lambda}\over 1+exp({r-c \over a})}\,\,,
\end{equation}
\\
where, $V_{0\lambda}= -28.0$ MeV, c=(1.128+0.439A$^{-2/3}$)A$^{1/3}$ and a=0.6
fm
give a very good account of the $B_{\Lambda}$ data. This procedure gives
$\Delta B_{\Lambda}$=0.40 MeV.

In the third approach, we consider a density-dependent $\Lambda$-nucleus
potential\cite{millener88,ahmad85,mian87} of the form

\begin{equation}
U_{\Lambda}(r)=A\rho(r)+B\rho^{4/3}(r),
\end{equation}
where $\rho (r)$ represents the nucleon density. These are taken from
Ref.\cite{vries87}.
We choose s- and p-orbits in $_{\,\,\Lambda}^{16}$O and the p- and f-orbits in
$_{\,\,\Lambda}^{89}$Y to fix the parameters A and B.  The experimental binding
energies are
12.5$\pm$0.35 and 2.5$\pm$0.5 MeV in $_{\,\,\Lambda}^{16}$O and 16.0$\pm$1.0
and 2.5$\pm$1.0
MeV in $_{\,\,\Lambda}^{89}$Y.  Fitting these energies results in a reasonable
fit to all
binding energies from $_{\,\,\Lambda}^{11}$C to $_{\,\,\Lambda}^{89}$Y for
$\ell_{\Lambda}\le 3$.
The resulting $\Delta B_{\Lambda}$ is 0.76 MeV.

If we combine the results of the above three approaches and at the same time
bear in mind
that the experimental uncertainty in the $B_{\Lambda}$ value of
$_{\,\,\Lambda}^{16}$O
is $\pm$0.35 MeV, we may reliably fix the empirical $B_{\Lambda}$ value of
$_{\,\,\Lambda}^{17}$O as 13.0 $\pm$ 0.4 MeV. We shall make use of this value
in our
calculations. Other approaches, such as relativistic mean-field theories or the
local-density approximation using a Skyrme interaction can also be employed in
the empirical
determination of the $B_{\Lambda}$ value of $_{\,\,\Lambda}^{17}$O. But, since
these
approaches are consistent with the approaches that we have adopted above, we do
not feel
that their inclusion will affect our results.

\section{Hamiltonian}

In an A-baryon hypernucleus, we will consider the first A-1 baryons to be
nucleons.
We will use $\Psi$ to refer to the full wavefunction of the hypernucleus
and $\Psi_{N}$ to refer to the ground-state wave function of the A-1 nucleons.
The full Hamiltonian H can be written as
\begin{equation}
H=H_{N}+H_{\Lambda}\, ,
\end{equation}
where $H_{N}$ is the nucleon Hamiltonian:
\begin{equation}
H_{N}=-\displaystyle\sum_{i=1}^{A-1} {{\hbar^{2}} \over {2m}}
{\mbox{\boldmath$\nabla$}}_{i}^{2}
 \ +\sum_{i<j}^{A-1}
v_{ij} \ +\sum_{i<j<k}^{A-1}V_{ijk}\, ,
\end{equation}
and
\begin{equation}
H_{\Lambda}=-{\hbar^{2} \over {2m_{\Lambda}}}
{\mbox{\boldmath$\nabla$}}_{\Lambda}^{2} \ +\sum_{i=1}^{A-1}
v_{i\Lambda} \ +\sum_{i<j}^{A-1}V_{ij\Lambda}\, ,
\end{equation}
is the part of the full Hamiltonian due to the $\Lambda$-particle.

The $\Lambda$-separation energy, $B_{\Lambda}$, of a hypernucleus is then given
by

\begin{equation}
B_\Lambda = {{<\Psi_{N}|H_{N}|\Psi_{N}>} \over {<\Psi_{N}|\Psi_{N}>}}- {{<\Psi
|H|\Psi >}
\over{<\Psi |\Psi >}}\, .
\end{equation}
Our goal is to calculate $B_{\Lambda}$ using a variational principle for the
two components
of Eq.(3.4). In this section we briefly describe the two- and three-body baryon
interactions
that we have used in this study.

\subsection{$\Lambda$N potential}

Two-pion-exchange(TPE) is a dominant part of the $\Lambda$N potential, that in
turn is
mainly determined by the strong tensor one-pion-exchange(OPE) component acting
twice.
Moreover, there is the K-exchange interaction that primarily contributes to the
$\Lambda$N
exchange potential. The tensor part of the $\Lambda$N interaction is very weak
because
the shorter range $\bar{K}$- and $K^{*}$-exchanges that are responsible for
this are of
opposite sign and nearly cancel each other. (In the case of the NN interaction
the
$\pi$-exchange and $\rho$-exchange tensor components do not cancel so
completely,
because their masses are quite different.)

We use an Urbana-type\cite{lagaris} potential with spin- and space-exchange
components and a
TPE tail which is consistent with $\Lambda p$ scattering below the $\Sigma$
threshold,

\begin{equation}
v_{\Lambda N}(r)=v_{0}(r)(1-\epsilon+\epsilon P_{x})+{1\over 4} v_{\sigma}
T_{\pi}^{2}(r){\mbox{\boldmath$\sigma$}}_{\Lambda}\cdot
{\mbox{\boldmath$\sigma$}}_{N},
\end{equation}

\begin{equation}
v_{0}(r)=v_{c}(r)-\bar{v} T_{\pi}^{2}(r)\,,
\end{equation}

\begin{equation}
v_{c}(r)={W_{c}\over 1+exp{\left({r-R\over a}\right)}}\,\,,
\end{equation}
where $T_{\pi}$(r) is the OPE tensor potential
\begin{equation}
T_{\pi}=\left({1+{3\over x }+{3\over x^{2}}}\right) {e^{-x}\over x}
\left({1-e^{-cr^{2}}}\right)^{2}\,,
\end{equation}
x=$\mu$r, $\mu$=0.7 fm$^{-1}$ is the pion mass,
and the cut-off parameter c=2.0 fm$^{-2}$. $P_{x}$ is the space exchange
operator
and $\epsilon$ is the corresponding exchange parameter.  The $\bar{v}\equiv
(v_{s}+3v_{t})/4$
and $v_{\sigma}\equiv v_{s}-v_{t}$ are respectively the spin-average and
spin-dependent
strengths, where $v_{s}$ and $v_{t}$ denote the singlet and triplet state
depths,
respectively.  (Note that following the convention of Ref.\cite{bodmer88}, the
Hamiltonian
effectively contains +$v_c$, --$\bar v$, +$v_{\sigma}$, --$v_s$, and --$v_t$.)
Finally, $v_{c}$(r) is a short range Saxon-Wood repulsive potential.
The various parameters are

\begin{equation}
v_{s}=6.33\,$MeV$,\,\, v_{t}=6.1\,$MeV$,\,\, \epsilon =0.3,\,\,
W_{c}=2137\,$MeV$,\,\, R=0.5\,$fm$,\,\, a=0.2\,$fm$\,.
\end{equation}
These parameters are consistent with the low energy $\Lambda p$ scattering
data that essentially determine the spin-average potential $\bar{v}$. The
parameter
$\epsilon$ for the space-exchange strength is fairly well determined from the
$\Lambda$
single-particle scattering data\cite{qusmani}.
For a detailed account of the determination of the other
parameters see Ref.\cite{bodmer88}.  We point out that because of the
non-central
$\Lambda$N and $\Lambda$NN correlations introduced in the next section, the
$\Lambda$N
spin-spin potential will have a non-zero contribution even in a closed-shell
system such
as $^{17}_{\,\,\Lambda}$O.

\subsection{$\Lambda$NN potential}

When a $\Lambda$N potential that fits the $\Lambda$p scattering is used, the
$B_{\Lambda}$
for hypernuclei
with A$\geq$5 are almost a factor of two too large. This is an old result that
has
been confirmed by various analyses. In the present work we also find that the
use of a
realistic NN interaction does not alleviate this over-binding problem. As in
the
previous studies, to resolve the over-binding problem, we incorporate a
three-body
$\Lambda$NN interaction.

We consider here two types of $\Lambda$NN potentials that arise from projecting
out
$\Sigma$,$\Delta$, etc. degrees of freedom from a coupled channel formalism.
These are
the dispersive and the TPE $\Lambda$NN potentials designated as $V_{\Lambda
NN}^{D}$
and V$_{\Lambda NN}^{2\pi}$, respectively (see Fig. 1). V$_{\Lambda NN}^{D}$,
is expected to be repulsive, and,
following Ref.\cite{bodmer88}, we assume the phenomenological form

\begin{equation}
V_{\Lambda NN}^{D}=W_{0}T_{\pi}^{2}(r_{1\Lambda})T_{\pi}^{2}(r_{2\Lambda}),
\end{equation}
where $W_{0}$ is the strength of the potential and $T_{\pi}(r_{i\Lambda})$ is
given by
Eq.(3.8). V$_{\Lambda NN}^{2\pi}$ consists of two parts corresponding to p and
s wave $\pi \Lambda$
interactions\cite{bhaduri67}
\begin{equation}
V_{\Lambda NN}^{2\pi}=W_{p}+W_{s},
\end{equation}
where
\begin{equation}
W_{p}=-\left({C_{p}\over 6}\right)
({\mbox{\boldmath$\tau$}}_{1}.{\mbox{\boldmath$\tau$}}_{2})\{X_{1\Lambda},X_{2\Lambda}\},
\end{equation}
\begin{equation}
W_{s}={C_{s}({\mbox{\boldmath$\tau$}}_{1}\cdot{\mbox{\boldmath$\tau$}}_{2})({\mbox{\boldmath$\sigma$}}_{1}\cdot{\bf r}_{1\Lambda})({\mbox{\boldmath$\sigma$}}_{2}
\cdot{\bf r}_{2\Lambda})(\mu r_{1\Lambda}+1)(\mu r_{2\Lambda}+1)Y(r_{1\Lambda})
Y(r_{2\Lambda})
\over (\mu r_{1\Lambda}r_{2\Lambda})^2},
\end{equation}
\begin{equation}
Y(r)={e^{-\mu r}\over \mu r} \left({1-e^{-cr^{2}}}\right),
\end{equation}
\begin{equation}
\{A,B\}=AB+BA,
\end{equation}
and
\begin{equation}
X_{i\Lambda}=({\mbox{\boldmath$\sigma$}}_{i} \cdot
{\mbox{\boldmath$\sigma$}}_{\Lambda})Y(r_{i\Lambda})+S_{i\Lambda}
T_{\pi}(r_{i\Lambda}).
\end{equation}
Here $X_{i\Lambda}$ is the one-pion exchange operator, and $S_{i\Lambda}$ is
the tensor operator.
The component $W_{s}$ is quite weak and, as
in previous studies, we neglect it here; we feel that its effect should be
studied in
future work.

There are theoretical as well as phenomenological estimates for $C_{p}$; but
for $W_{0}$
the estimates are purely phenomenological. For example, for $W_{0}\approx$
0.02, the
reduction in the $\Lambda$-binding to nuclear matter(using central
correlations) is
approximately in accord with the suppression obtained in coupled
channel ($\Lambda N \rightarrow \Sigma N$) reaction matrix calculations. For
$C_{p}$
theoretical estimates give 1$\sim$2 MeV; however, the phenomenological values
may not lie
in this region as the results depend sensitively on the cutoff parameter c that
appears
in Eq.\ (3.8). In the present study, we have obtained results as a function of
the
values of these parameters.

\subsection{Two- and three-nucleon potentials}
For the nuclear part of the Hamiltonian, we use NN and NNN potentials that have
been
previously used to study various nuclei, including $^{16}$O\cite{pieper92}.
The NN potential contains
the first six terms of the
Argonne $v_{14}$\cite{wiringa84} potential and a Coulomb term:
\begin{equation}
v_{ij}=\sum_{p=1}^{6}v^{p}(r_{ij})O_{ij}^{p}+v_{Coul}(r_{ij}) O^{Coul}_{ij}
\end{equation}
where the operators are
\begin{equation}
\left.
\begin{array}{lcl}
O_{ij}^{p=1,6} &=& 1,{\mbox{\boldmath$\tau$}}_{i} \cdot
{\mbox{\boldmath$\tau$}}_{j},{\mbox{\boldmath$\sigma$}}_{i} \cdot
{\mbox{\boldmath$\sigma$}}_{j},
\left({\mbox{\boldmath$\sigma$}}_{i} \cdot
{\mbox{\boldmath$\sigma$}}_{j}\right)
\left({\mbox{\boldmath$\tau$}}_{i} \cdot {\mbox{\boldmath$\tau$}}_{j}\right)
,S_{ij}, S_{ij}\left({\mbox{\boldmath$\tau$}}_{i} \cdot
{\mbox{\boldmath$\tau$}}_{j}\right)\,,
\\
O^{Coul}_{ij} &=& \frac{1}{4} (1+\tau_{z,i}) (1+\tau_{z,j})\,.
\end{array}
\right\}
\end{equation}
We shall also refer to these operators by the abbreviations c, $\tau, \,
\sigma,\,
\sigma\tau$, t, and t$\tau$.  In PWP it was found that the 7 $\leq$ $p$ $\leq$
14
terms of Argonne $v_{14}$ and the corresponding $p$ = 7,8 correlation operators
gave a
net contribution of only -0.45 MeV/nucleon.  We assume that the presence of a
$\Lambda$
will not significantly modify this result and hence we can safely omit all
potential
and correlation operators for $p$ $\geq$ 7 when computing
$B_{\Lambda}$($^{17}_{\,\,\Lambda}$O).

The NNN potential is of the Urbana type, which consists of dispersive and
two-pion-exchange terms:

\begin{equation}
V_{ijk}=V_{ijk}^{D}+V_{ijk}^{2\pi},\\
\end{equation}
\begin{equation}
V_{ijk}^{D}=\sum_{cyc}U_{0}T_{\pi}^{2}(r_{ij})T_{\pi}^{2}(r_{jk}),\\
\end{equation}
\begin{equation}
V_{ijk}^{2\pi}=\sum_{cyc}A_{0}\left(
\{X_{ij},X_{jk}\}\{{\mbox{\boldmath$\tau$}}_{i}.{\mbox{\boldmath$\tau$}}_{j},{\mbox{\boldmath$\tau$}}_{j}.{\mbox{\boldmath$\tau$}}_{k}\}
+{1\over
4}[X_{ij},X_{jk}][{\mbox{\boldmath$\tau$}}_{i}.{\mbox{\boldmath$\tau$}}_{j},{\mbox{\boldmath$\tau$}}_{j}.{\mbox{\boldmath$\tau$}}_{k}]\right),
\end{equation}
where the square brackets represent the commutator [A,B]=AB-BA. The constants
$A_{0}$ and $U_{0}$ have the values -0.0333 and 0.0038 in Urbana model
VII\cite{schiavilla},
which we use here.

\section{the variational wave functions}
We assume that a good variational wavefunction for a hypernucleus with a
closed-shell
nuclear core and a $\Lambda$-particle can be written as
\begin{equation}
|\Psi >  =
\left[\prod_{IT}(1+U_{ij\Lambda}+U_{ijk})\right]
\left[\prod_{i=1}^{A-1} {(1+U_{i\Lambda})}\right] \ \left[S\prod_{i<j}^{A-1}
{(1+U_{ij})}\right] |\Psi_{J}>,
\end{equation}
\begin{equation}
|\Psi_{J}>=\prod_{i=1}^{A-1} f_{c}^{\Lambda}(r_{i\Lambda})\prod_{i<j}^{A-1}
f_{c}(r_{ij})\phi_{\Lambda}(r_{\Lambda}){\bf A} |\Phi^{A-1}>.
\end{equation}
Here $U_{ijk}$ represents a three-baryon correlation operator that has the same
structure as $V_{ijk}$,
\begin{equation}
       U_{ijk} = \delta \tilde{V}_{ijk},
\end{equation}
\begin{equation}
       U_{ij\Lambda} = \delta_{\Lambda} \tilde{V}_{ij\Lambda} \, .
\end{equation}
The $\tilde{V}_{ijk}$ differs from $V_{ijk}$  through the range c of the cutoff
functions
of $Y_{\pi}$(r) and $T_{\pi}$(r).  The parameter $\delta$ is referred to as
$\epsilon$ in
PWP; we use $\delta$ here to avoid confusion with the $\epsilon$ of
the space-exchange potential in Eq.\ (3.5).
The parameters $\delta,\,\delta_{\Lambda}$, c and
c$_{\Lambda}$
are determined variationally.  The label IT stands for independent triplet
product
of 1+$U_{ijk}$. Thus,

\begin{eqnarray}
\prod_{IT}\left(1+U_{ijk}+U_{ij\Lambda}\right)&=&1+\sum_{i<j}U_{ij\Lambda}+
\sum_{^{\,\,\,\,\,\,\,\,\,\,\,\,i<j}_{i'<j'<k'\neq i,j}}
U_{ij\Lambda}
U_{i^{'}j^{'}k^{'}}+\sum_{i<j<k}U_{ijk}
\nonumber \\
&+& \sum_{^{\,\,\,\,\,\,\,\,\,\,\,i<j<k}_{i^{'}<j^{'}<k^{'}\neq i,j,k}}
U_{ijk} U_{i^{'}j^{'}k^{'}}+\cdots
\end{eqnarray}
The neglected terms are of the type $U_{ijk}U_{i^{'}j^{'}k}$. This restriction
makes
the three-body correlations much simpler to use.  As is discussed in PWP, the
$U_{ijk}$
and $U_{ij\Lambda}$
should ideally act last as in Eq.\ (4.1). However, this requires considerably
more
computer time. The improvement in the energy of $^{16}$O obtained by this was
found to
be only -0.19(7) MeV/nucleon. In the present work we ignore this correction and
compute the energies of both $_{\,\,\Lambda}^{17}$O and $^{16}$O with the
three-baryon
correlations acting first.

Each operator in the two-baryon interaction can induce the corresponding
correlation.
The $f_{c}^{\Lambda}$ and the $f_{c}$ are central correlations that are
primarily
generated by
the repulsive cores in the two-baryon interactions. For $U_{ij}$ and
$U_{i\Lambda}$,
we make the following choice

\begin{equation}
U_{ij}=\sum_{p=2}^{n} \beta_{p} u_{p}(r_{ij})O_{ij}^{p},
\end{equation}
and
\begin{equation}
U_{i\Lambda}=\sum_{p=2}^{m} u_{p}^{\Lambda}(r_{i\Lambda})O_{i\Lambda}^{p}.
\end{equation}
The notation $S\prod$ in Eq.\ (4.1) represents a symmetrized-product of the
non-commuting
operators $U_{ij}U_{jk}\cdots$. Previous studies\cite{pandharipande79,pieper92}
on
few-body nuclei
and $^{16}$O demonstrate that it is probably sufficient to use 2$\leq$ p $\leq$
6 in
Eq. (4.6).
The pair correlation functions $f_{c}$ and $u_{p}$ are generated by minimizing
the
two-body cluster energy using a quenched potential:
\begin{equation}
\tilde{v}_{ij}=\sum_{p=2}^{6} \alpha v_{p}(r_{ij})O_{ij}^{p}.
\end{equation}
The two-body cluster contribution has been minimized for infinite nuclear
matter at
Fermi momentum $k_{F}$, with the boundary conditions $f_{c}(r>d)=1$, and
$u_{p}(r>d)=0$,
for $p=\tau,\sigma,$ and $\sigma \tau$, and $u_{p}(r>d_{t})=0$, for $p=t$ and
$t\tau$ with their first derivatives zero at $r=d$ or $d_{t}$.

For the $U_{i\Lambda}$ we consider

\begin{equation}
 U_{i\Lambda}=u_{\sigma}^{\Lambda}(r_{i\Lambda})
{\mbox{\boldmath$\sigma$}}_{\Lambda}
\cdot {\mbox{\boldmath$\sigma$}}_{i}+ u_{P_{x}}(r_{i\Lambda})P_{x}.
\end{equation}
In the present study, we have omitted the second, i.e., the exchange
correlation term in
Eq.(4.9). Inclusion of this term increases the computation effort by several
fold
and preliminary results indicate that it gives a small contribution. This will
be the
subject of a future publication. The spin correlation is

\begin{equation}
u_{\sigma}^{\Lambda}={{f_{t}^{\Lambda}-f_{s}^{\Lambda}}\over f_{c}^{\Lambda}},
\end{equation}
where $f_{s}^{\Lambda}$ and $f_{t}^{\Lambda}$ are the solutions of Schroedinger
equations
with quenched $\Lambda$N potentials in singlet and triplet states respectively:

\begin{equation}
\left[{-{\hbar^{2}\over 2\mu_{\Lambda N}}{\bf \nabla}^{2} +\tilde{v}_{s(t)}
(r_{\Lambda N}) }\right] f^{\Lambda}_{s(t)} = 0\, .
\end{equation}
The potentials $\tilde{v}_{s(t)}$ are quenched in the two-pion and
spin-exchange
parts of the central and spin channels:
\begin{equation}
\tilde v_s (r) = v_c (r) - (\alpha_{2\pi}\, \bar v +
\case{3}{4}\,\alpha_{\sigma}
v_{\sigma}) T^{2}_{\pi} (r)\, ,
\end{equation}
\begin{equation}
\tilde v_t (r) = v_c (r) - (\alpha_{2\pi}\, \bar v - \case{1}{4}\,
\alpha_{\sigma}
v_{\sigma}) T_{\pi}^{2} (r)\, .
\end{equation}
The  spin-averaged correlation function $f_{c}^{\Lambda}$ is given by:
\begin{equation}
f_{c}^{\Lambda}={{f_{s}^{\Lambda}+3f_{t}^{\Lambda}}\over 4}.
\end{equation}
The $f_{s}^{\Lambda}$,$f_{t}^{\Lambda}$ and $f_{c}^{\Lambda}$ have been
obtained
by minimizing the two-body cluster energy for the $\Lambda$-binding to nuclear
matter
with the asymptotic condition $f_{c}^{\Lambda}(r>d_{\Lambda})=1$.

The $\phi_{\Lambda}$ represents a bound-state wavefunction of a
$\Lambda$-particle of
mass m$_{\Lambda}$ moving in a Woods-Saxon potential that is bound to a nucleus
of
mass (A-1)m:
\begin{equation}
V_{\Lambda}(r_{\Lambda})={V_{\Lambda}\over 1+exp{\left({r_{\Lambda}-
R_{\Lambda}\over a_{\Lambda}}\right)}}.
\end{equation}
The parameters $V_{\Lambda},R_{\Lambda}$ and $a_{\Lambda}$ are determined
variationally.
The Slater determinant ${\bf A}\left|\Phi^{A-1}\right>$ consists of orbitals of
nucleons
of mass m bound to a hypernucleus of mass (A-2)m+m$_{\Lambda}$ moving in a
Woods-Saxon
wine-bottle potential
\begin{equation}
V(r)=V_{s}\left({{1\over 1+e^{\left(r-R_{s})/a_{s}\right)}}-\alpha_{s}
e^{-\left(r/\rho_{s}\right)}}\right),
\end{equation}
with $V_{s}, R_{s},a_{s},\alpha_{s},$ and $\rho_{s}$ as variational parameters.
The coordinates of all the one-body orbitals are measured from the center of
mass
of the whole system, thus making $\Psi_{J}$ and $\Psi$ translationally
invariant, i.e.,
\begin{equation}
\tilde{\bf r}_{i}={\bf r}_{i}-{\bf R}_{c.m.}\, ,\\
\end{equation}
\begin{equation}
{\bf R}_{c.m.}={m{\sum_{i=1}^{A-1}}{\bf r}_{i}+m_{\Lambda} {\bf
r}_{\Lambda}\over m(A-1)+
m_{\Lambda}} \, .
\end{equation}

\section{the cluster expansion}

We briefly outline the general framework for the cluster expansion of PWP
to calculate the expectation values of various operators.  These expectation
values
are needed in the evaluation of the energy using the variational wave function
(4.1).  We
demonstrate the cluster expansion for the two-body NN and $\Lambda$N
potentials:

\begin{equation}
{\left< \Psi |\sum v_{ij}+ \sum v_{i\Lambda}|\Psi \right>\over
{\left< \Psi|\Psi \right>}} = {N\over D} \,\, .
\end{equation}
The $N$ and $D$ can be expanded as a sum of n-body contributions
\begin{equation}
N=\sum^{A-1}_{i<j} n_{ij} + \sum^{A-1}_{i=1} n_{i\Lambda} + \sum^{A-1}_{i<j<k}
n_{ijk}
+\sum^{A-1}_{i<j} n_{ij\Lambda} + \cdots \, ,
\end{equation}
\begin{equation}
D=1+\sum^{A-1}_{i<j} d_{ij} + \sum^{A-1}_{i=1} d_{i\Lambda} + \cdots \, .
\end{equation}

The expressions for the purely nuclear $n_{ij}$, $d_{ij}$, $n_{ijk}$, etc., may
be found in
PWP.  The contributions of clusters containing a $\Lambda$ are similar, e.g.\
\begin{equation}
n_{i\Lambda} = \left<(1+U_{i\Lambda})^{\dag}\,v_{i\Lambda}
(1+U_{i\Lambda})\right> \, ,
\end{equation}

\begin{equation}
d_{i\Lambda}=\left<(1+U_{i\Lambda})^{\dag}\,(1+U_{i\Lambda})\right>-1 \, ,
\end{equation}

\begin{eqnarray}
n_{ij\Lambda}&=&\left<\right. (1+U^{\dag}_{ij\Lambda}) (1+U^{\dag}_{ij})
P\left[(1+U^{\dag}_{i\Lambda})
(1+U^{\dag}_{j\Lambda})\right] \left[v_{ij} + v_{i\Lambda} +
v_{j\Lambda}\right] \nonumber\\
					&&   \times P'\left[(1+U_{i\Lambda}) (1+U_{j\Lambda})\right] (1+U_{ij})
(1+U_{ij\Lambda}) \left.\right> - n_{ij} - n_{i\Lambda} - n_{j\Lambda} \, ,
\end{eqnarray}
where $P$ and $P'$ are permutation operators.  In these expressions

\begin{equation}
\left<\theta\right> = {\left<\Psi_J |\theta |\Psi_{J'}\right> \over
\left<\Psi_{J'}
| \Psi_{J'}\right>} \, ,
\end{equation}
where $\Psi_{J'}$ denotes the $\Psi_{J}$ of Eq. (4.2) without the
anti-symmetrization
operator.

The expansions (5.2) and (5.3) for $N$ and $D$ are divergent\cite{pieper92}.
We obtain a convergent linked-cluster expansion by expressing
\begin{equation}
{N \over D} = \sum^{A-1}_{i<j} c_{ij} + \sum^{A-1}_{i=1} c_{i\Lambda} + \cdots
\end{equation}
The various $c_{ij}$ and $c_{i\Lambda}$ etc.\ are obtained from the equation

\begin{equation}
N=\left[\sum^{A-1}_{i<j} c_{ij} + \sum^{A-1}_{i=1} c_{i\Lambda} + \cdots
\right] D \, ,
\end{equation}
by equating terms with the same $ij$, $i\Lambda$, $ijk$, $ij\Lambda$, etc.
Thus,
\begin{equation}
c_{i\Lambda} = {n_{i\Lambda} \over 1+d_{i\Lambda}} \, ,
\end{equation}
and
\begin{equation}
c_{ij\Lambda} = {n_{ij\Lambda} - c_{ij}
(d_{j\Lambda} + d_{i\Lambda}) - c_{i\Lambda}
(d_{ij} + d_{j\Lambda}) - c_{j\Lambda}
(d_{ij} + d_{i\Lambda}) -
(c_{ij} + c_{i\Lambda} + c_{j\Lambda}) d_{ij\Lambda} \over
1+d_{ij} + d_{i\Lambda} + d_{j\Lambda} + d_{ij\Lambda}}\,.
\end{equation}

In the present work, we have used the CEA expansion of PWP so that all clusters
of a given
spin, isospin, and $\Lambda$ content are averaged together.

\section{results}
\subsection{Variational parameters}
We made detailed variational parameter searches for two cases: 1) with no
$\Lambda$NN
potential and hence no $U_{ij\Lambda}$ correlation, and 2) using a $\Lambda$NN
potential
with $C_p$ = 0.7 MeV and $W_0$ = 0.015 MeV.  The rest of the Hamiltonian was as
described in Sec. III for both cases; in particular the NNN potential and
$U_{ijk}$
correlation were used in both cases.

For the case with no $\Lambda$NN potential, we found that the optimal values of
all of the
nucleon correlation parameters are the same as was found in PWP for
$^{16}$O, except that the well depth, $V_s$, of the Woods-Saxon potential
changes from -49.1
MeV to -48.9 MeV to maintain the same p-wave separation energy, 14.0 MeV, with
the 17-body
instead of 16-body reduced mass.  The reader is referred to PWP for these
parameter values.  The optimal parameters for correlation terms involving the
$\Lambda$
are: for the $\Lambda$ Woods-Saxon well,

\begin{equation}
V_{\Lambda}= -28.3 \,\,$MeV$;\,\,\,\,\,\,R_{\Lambda} = 3.2 \,\,$fm$;\,\,\,\,\,\
a_{\Lambda} = 0.5\,\, $fm$\,,
\end{equation}
which gives an s-wave separation energy of 15.0 MeV; for the $U_{i\Lambda}$

\begin{equation}
\alpha_{2\pi} = \alpha_{\sigma} = 1.0\,;\,\,\,\,\,\, d_{\Lambda} = 2.8
\,\,$fm$\,.
\end{equation}

In the presence of the $\Lambda$NN potential with $C_p$ = 0.7 MeV, $W_0$ =
0.015 MeV,
we found that only two of the above optimal values had to be changed.  These
are the
quenching parameters $\alpha$ in the $NN$ correlation and $\alpha_{2\pi}$ in
the
$N\Lambda$ correlation.  The variational energy is sensitive to $\alpha$ and we
made
several searches at other values of $C_p$ to determine

\begin{equation}
\alpha = 0.94 - 0.1\, C_p\,,\,\,\,\,\,\ 0 \leq C_p \leq 1.2 \,\, $MeV$\,.
\end{equation}
The sensitivity to $\alpha_{2\pi}$ is weak and we used

\begin{equation}
\alpha_{2\pi} = 0.95 \,,\,\,\,\,\,\, 0.7 \leq C_p \leq 1.2 \,\,$MeV$\,.
\end{equation}
In addition to these parameters we found  for the $U_{NN\Lambda}$

\begin{equation}
\delta_{\Lambda} = -0.0013\,; \,\,\,\,\,\, c_{\Lambda} = 1.6 \,\,$fm$^{-2}\,,
\end{equation}
for all $C_p$ and $W_0$ considered.

\subsection{Variational energies}
Tables I and II show various components of the $^{17}_{\,\,\Lambda}$O energy
for the cases of
no $\Lambda$NN potential and $C_p$ = 0.7 MeV, $W_0$ = 0.015 respectively.  The
cluster
expansion for terms involving the $\Lambda$ is converging well and it appears
that it is
not necessary to extrapolate beyond the four-body clusters for these terms.  To
get an
accurate total energy of the $^{17}_{\,\,\Lambda}$O nucleus, it would be
necessary to
extrapolate the values of $V_{NNN}$ as was done in PWP.  However since we
are mainly interested in $B_{\Lambda}$, we do not do that here and instead
subtract
an unextrapolated $^{16}$O energy.

One important result shown in Table II is that the expectation of
$V_{\Lambda}^{2\pi}$ is
substantial and negative.  This arises from the non-central correlations in the
wave
function.  A purely Jastrow wave function ($U_{ij} = U_{ijk} = U_{i\Lambda} =
U_{ij\Lambda} = 0$) gives $\left<V^{2\pi}_{\Lambda NN}\right>$  = 0.9 MeV;
including $U_{ij}$ (with $\alpha$ = 0.94), $U_{ijk}$, and $U_{i\Lambda}$
(with $\alpha_{2\pi}$ = 1.0),
but no $U_{ij\Lambda}$, results in $\left<V^{2\pi}_{\Lambda NN}\right>$ = -4.2
MeV.
Including $U_{ij\Lambda}$ also but still keeping $\alpha$ = 0.94,
$\alpha_{2\pi}$ = 1.0,
gives $\left<V^{2\pi}_{\Lambda NN}\right>$ = -7.9 MeV, while lowering $\alpha$
to
the optimal value
of 0.87 and $\alpha_{2\pi}$ to 0.95 reduces this to -5.0 MeV (this loss of
binding is offset
by changes in the expectation values of other parts of the Hamiltonian).  These
results can
be understood as follows: by using the relation

\begin{equation}
{{\mbox{\boldmath$\sigma$}}} \cdot \bf{A}\, {{\mbox{\boldmath$\sigma$}}} \cdot
\bf{B} = \bf{A} \cdot \bf{B}
+ i\,{{\mbox{\boldmath$\sigma$}}} \cdot (\bf{A} \times \bf{B})\,,
\end{equation}
the $\{ X_{1\Lambda},X_{2\Lambda}\}$ appearing in
$V^{2\pi}_{\Lambda NN}$ can be expressed in terms of the operators
${{\mbox{\boldmath$\sigma$}}}_{1} \cdot {\bf{r}}_{1\Lambda}\,
{{\mbox{\boldmath$\sigma$}}}_{2} \cdot {\bf{r}}_{2\Lambda}$,
${{\mbox{\boldmath$\sigma$}}}_{1} \cdot {\bf{r}}_{1\Lambda}\,
{{\mbox{\boldmath$\sigma$}}}_{2} \cdot {\bf{r}}_{1\Lambda}$,
${{\mbox{\boldmath$\sigma$}}}_{1} \cdot {\bf
r}_{2\Lambda}\,{{\mbox{\boldmath$\sigma$}}}_{2} \cdot {\bf r}_{2\Lambda}$,
and ${{\mbox{\boldmath$\sigma$}}}_{1} \cdot {{\mbox{\boldmath$\sigma$}}}_{2}$
and hence is a generalization of the tensor
operator $S_{12}$.  However the expectation value of $S_{12}$ in a Jastrow wave
function
for a closed-shell nuclear system
is zero, while the expectation value of $S^{2}_{12}$ is non-zero.  Hence the
$S_{12}$
operator in $U_{12}$ completely changes $\left<V^{2\pi}_{\Lambda 12}\right>$.
Of course the
$U_{ij\Lambda}$ further enhances its contribution.

The $^{16}$O energy that corresponds to the present calculation ($U_{ijk}$ acts
first, use
of only the first six operators in Argonne $v_{14}$, and no extrapolation) is
-101.0(9) MeV.
We emphasize that, because of the above approximations, this energy is to be
used only
in comparison with the $^{17}_{\,\,\Lambda}$O energies.  The resulting
$B_{\Lambda}$($^{17}_{\,\,\Lambda}$O) are 27.5(2.0) MeV for no $V_{\Lambda NN}$
and 13.5(1.8)
MeV for $C_p$ = 0.7, $W_0$ = 0.015, which are to be compared with the empirical
value
of 13.0(4) found in Sec.\ II.  Thus even with realistic NN potentials and
correlations,
$^{17}_{\,\,\Lambda}$O is very overbound if no $V_{\Lambda NN}$ is used.
However a reasonable
$V_{\Lambda NN}$ results in a $B_{\Lambda}$ consistent with the empirical
value.

To study the dependence of $B_{\Lambda}$ on the strength of the $V_{\Lambda
NN}$, we made
a number of calculations with different values of $C_p$ and $W_0$.  In each
case the NN
quenching parameter was chosen according to Eq.\ (6.3).  To minimize
statistical errors,
we made correlated difference calculations using the $C_p$ = 0.7,
$W_0$ = 0.015 random walk.  Table III presents the resulting changes,
$\delta B_{\Lambda}$, in $B_{\Lambda}$.  All of
these values of $B_{\Lambda}$ and $\delta B_{\Lambda}$ are well fit by the
formula

\begin{equation}
B_{\Lambda} = 27.3 - 8.9\,C_p + 11.2\,C^{2}_{p} - 870.\,W_0\,;
\end{equation}
the statistical error of $B_{\Lambda}$ is $\pm$ 1.6 MeV.
The quadratic dependence on $C_p$ comes from $U_{ij\Lambda}$; the contribution
of the
dispersive term in $U_{ij\Lambda}$ (and also in $U_{ijk}$) is very small.  When
comparable calculations of other hypernuclei are available, Eq.\ (6.7) and the
empirical
value of $B_{\Lambda}$($^{17}_{\,\,\Lambda}$O) = 13.0(4) can be used to
uniquely determine the
values of $C_p$ and $W_0$ (or to show that a different Hamiltonian is needed if
no fit
can be found).

\subsection{Densities and polarization of $^{16}$O core}
Figure 2 shows point nucleon and $\Lambda$ densities for the calculations of
Tables I
and II.  The density of $^{16}$O is also shown.  For the case with no
$V_{\Lambda NN}$, the nuclear correlation parameters were not
changed from those used in $^{16}$O.  The resulting $^{16}$O density is,
however,
reduced near the origin and somewhat more peaked at $r$ = 1.4 fm.  This is
presumably due to the repulsive $f_{\Lambda N}$ which pushes the nucleons away
from the
$\Lambda$ which is strongly localized near the origin.

With $V_{\Lambda NN}$, the NN quenching parameter was significantly reduced.
This
results in a slightly more repulsive $f_c$ ($\alpha$ does not quench the
central
part of $V_{NN}$) and so the nuclear density is reduced for $r$
$\tiny^{<}_{\sim}$
2.2 fm.  The nuclear kinetic and potential energies (see Tables I and II) for
the no
$V_{\Lambda NN}$ case are separately larger in magnitude due to the higher
density.
It is probably accidental that the total nucleon energies for the two cases
are so nearly the same:
-90(2) MeV.  This value is 11 MeV less than the corresponding binding
energy of $^{16}$O, showing
that the $\Lambda$ significantly reduces the binding of the nucleons.

The density profile of the $\Lambda$ for the two cases is also shown in Fig.\
2, along
with the density corresponding to the one-body part of $\Psi$, i.e.
$|\phi_{\Lambda} (r)|$$^2$/4$\pi$.  In both cases the Jastrow part of $\Psi$,
i.e.
$\Pi f^{\Lambda}_{c} (r_{i\Lambda})$, significantly increases the $\Lambda$
density at the
origin.  This is presumably because $f^{\Lambda}_{c}$ is small for
$r_{i\Lambda}\rightarrow$ 0
and hence pushes the $\Lambda$ away from the high nuclear density around $r$ =
1.4 fm.
Because this density is larger for $V_{\Lambda NN}$ = 0, the central $\Lambda$
density
is also larger in this case.

\section{Conclusions}
In this study, we have extended the cluster Monte Carlo technique developed in
PWP for
$^{16}$O to $_{\,\,\Lambda}^{17}$O.  The cluster contributions that involve the
$\Lambda$ seem
to converge well. It thus seems sufficient to include terms up to four-body
clusters in the
calculation. These calculations have been performed for a number of sets of
$W_{0}$ and
$C_{p}$ which will be helpful in determining the parameters of the three-body
interaction
$V_{ij\Lambda}$
by fitting the $B_{\Lambda}$ values of $_{\,\,\Lambda}^{17}$O and other
hypernuclei. The present
calculations show that the use of non-central NN, NNN, N$\Lambda$ and
NN$\Lambda$ correlations completely change
the expectation value of the three-baryon $\Lambda$NN interaction found with
central
wave functions, and thus have a strong effect on the
choice of the parameters $W_{0}$ and $C_{p}$.  With such correlations,
reasonable values
of $C_p$ and $W_0$ give the correct $B_{\Lambda}$($^{17}_{\,\,\Lambda}$O).  We
also
find that the $\Lambda$ significantly changes the density profile and
energy of the 16 nucleons in $^{17}_{\,\,\Lambda}$O; the $\Lambda$NN
potential is particularly significant in this regard.

\acknowledgements
We thank R.\ B.\ Wiringa for many useful discussions.
This work was supported by the Indo-US joint collaboration program for which
the authors
acknowledge NSF grant INT-9011046.  This work was also supported in part by the
U.S.
Department of Energy, Nuclear Physics Division, under contract No.
W-31-109-ENG-38.
The calculations were made possible by a grant of computer time at the National
Energy
Research Supercomputer Center, Livermore, California. AAU is also grateful to
Professor A.
Zichichi, President, ICSC-World Laboratory, Lausanne, Switzerland for a
fellowship,
Professor S. Fantoni, Director, Interdisciplinary Laboratory, SISSA,
34014-Trieste,
Italy for extending the facilities to complete this work, Professor
Basheeruddin Ahmad, Vice Chancellor, Jamia Millia
Islamia, New Delhi for a grant of leave and support, and Professor Abdus Salam,
Director,
International Center for Theoretical Physics, 34014 Trieste, Italy for inviting
him to
ICTP where part of the work was carried out.  QNU acknowledges a DRS grant of
the Department
of Physics, Jamia Millia Islamia, New Delhi.  SCP thanks Professor Q.N.\ Usmani
and his
wife, Tasneem, for their generous hospitality while he was in New Delhi, and
Professor
S.\ Fantoni for an invitation to SISSA, where part of this paper was written.

\newpage
\begin{figure}
\caption{Terms contributing to $V^{D}_{\Lambda NN}$ and $V^{2\pi}_{\Lambda
NN}$.}
\end{figure}

\begin{figure}
\caption{One-body densities for the nucleons and $\Lambda$ in
$^{17}_{\,\,\Lambda}$O,
and for $^{16}$O.  The short-dashed curve is the
density corresponding to just $\phi_{\Lambda}$.}
\end{figure}

\newpage
\begin{table}[t]
\caption{Variational energies for $C_p$ = $W_0$ = 0.  All values are in
MeV.\,\,
Numbers in parentheses are statistical errors in the last digit.}
\begin{tabular}{lcddddd}
\hline   \\
 Clusters  && one-body & two-body     &  three-body      &  four-body	  &
Total\\
\\
\hline \\
Kinetic Energy$^{\rm a}$	&	&20.2(6)	&0.1(1)	&1.1(5)	&-0.7(3)	&20.7(9) \\
$\Lambda$N Potential	&$v_{0}(r)(1-\epsilon)$  &
															& 	-44.9(9)	&3.7(3)		&1.4(7)		&-47.3(12) \\
	& $v_{0}(r)\epsilon P_{x}$	&&-13.4(4)	&0.4(1)	&0.7(3)	&-12.2(5) \\
&$\frac{1}{4} v_{\sigma} T^{2}_{\pi}  {\mbox{\boldmath$\sigma$}}_{\Lambda}
\cdot {\mbox{\boldmath$\sigma$}}_{N}$ &
&0.34(3)	&-0.06(1)	&-0.05(2)	&0.22(3) \\
$\Lambda$ Energy		&&20.2(6)	&-57.9(11)	&-2.3(7)	&1.4(11)	&-38.6(10) \\
Nuclear Kinetic		&&317.(2)	&269.(2)	&-19.(3)	&11.(3)	&556.(5) \\
$NN$ Potential	&$v_{ij}$  &&-737.(4)	&111.(2)	&4.(4)	&-623.(5) \\
$NNN$ Potential &$V_{ijk}$  &&&-59.8(8)	&35.8(8)	&-23.9(8) \\
Nuclear Energy	&&317.(2)	&-468.(2)	&32.(2)	&29.(3)	&-90.(2) \\
Total Energy	&&337.(2)	&-526.(3)	&30.(3)	&30.(1)	&-128.5(18) \\
\\
\hline
\end{tabular}
\tablenotetext[1]{Includes nucleon kinetic energy from $\Lambda$N
correlations.}
\end{table}

\newpage
\begin{table}[t]
\caption{Variational Energies for $C_p$ = 0.7 MeV, $W_0$ = 0.015 MeV.  All
values are in
MeV.  Numbers in parentheses are statistical errors in the last digit.}
\begin{tabular}{lcddddd}
\hline   \\
Clusters  &        & one-body   & two-body   &  three-body           &
four-body			& Total \\
\\
\hline \\
$\Lambda$ Kinetic Energy$^{\rm a}$
&           &17.1(5)	&0.1(1)	&2.4(5)	&-1.0(4)	&18.6(6) \\
$\Lambda$N Potential	&$v_{0}(r)(1-\epsilon)$
											&&-39.5(7)	&-2.0(2)	&0.6(5)	&-40.9(8) \\
&$v_{0}(r)\epsilon P_{x}$	&&-12.5(3)	&0.5(1)	&-0.2(4)	&-12.2(4) \\
&$\frac{1}{4} v_{\sigma} T^{2}_{\pi} {\mbox{\boldmath$\sigma$}}_{\Lambda} \cdot
{\mbox{\boldmath$\sigma$}}_{N}$		&
&0.26(2)	&-0.06(1)	&0.00(2)	&0.20(2) \\
$\Lambda$NN Potential	&$V^{D}_{\Lambda NN}$
						&&&		14.1(4)	&-0.1(1)	&14.0(4) \\
& $V^{2\pi}_{\Lambda NN}$	&&& -6.5(3)	&1.5(2)	&-5.0(3) \\
$\Lambda$ Energy	&&17.1(5)	&-51.7(10)	&8.4(6)	&0.9(9)	&-25.3(7) \\
Nuclear Kinetic	&&309.(2)	&229.(2)	&-7.(2)	&-12.(2)	&520.(4) \\
$NN$ Potential	&$v_{ij}$  &&-682.(3)	&85.(2)	&10.(3)	&-587.(4) \\
$NNN$ Potential	&$V_{ijk}$  &&&-50.2(6)	&27.9(6)	&-22.3(7) \\
Nuclear Energy &&309.(2)	&-453.(2)	&28.(2)	&27.(2)	&-90.(2) \\
Total Energy	&&326.(2)	&-505.(2)	&36.(2)	&38.(2)	&-114.5(16) \\
\\
\hline
\end{tabular}
\tablenotetext[1]{Includes nucleon kinetic energy from $\Lambda$ correlations.}
\end{table}

\newpage
\begin{table}[t]
\caption{Differences $B_{\Lambda}(C_p,W_0)$-$B_{\Lambda}(C_p = 0.7, W_0 =
0.015)$.  The
NN quenching parameter, $\alpha$, is also given.}
\begin{tabular}{lddc}
\hline   \\
$C_p$				&	$W_0$				&	$\alpha$				&	$\delta B_{\Lambda}$ MeV \\
\\
\hline \\
0.7					       &0.01					             &0.87					            &$+4.4 \pm .3$
      \\
0.7					       &0.017					            &0.87					            &$-1.7 \pm .1$
      \\
0.7					       &0.02					             &0.87					            &$-4.4 \pm .3$
      \\
0.9					       &0.01					             &0.85					            &$+6.3 \pm .7$
      \\
0.9					       &0.015					            &0.85					            &$+1.9 \pm .5$
      \\
0.9					       &0.02					             &0.85					            &$-2.4 \pm .5$
      \\
1.0					       &0.015					            &0.84					            &$+2.8 \pm .8$
      \\
1.0					       &0.02					             &0.84					            &$-1.5 \pm .7$
      \\
\\
\hline
\end{tabular}
\end{table}

\end{document}